\documentclass[doublecol]{epl2}
\usepackage{graphicx}
\usepackage{amsmath}
\usepackage{amssymb}
\def\slaninafigdir{.}
\title{%
Interaction of molecular motors can enhance their efficiency
}%%
\shorttitle{%
Interaction of molecular motors
}%
\author{%
Franti\v{s}ek Slanina
\inst{1}%
\shortauthor{%
Franti\v{s}ek Slanina
}
}
\institute{
\inst{1}%
Institute of Physics,
 Academy of Sciences of the Czech Republic,
 Na~Slovance~2, CZ-18221~Praha,
Czech Republic
}
\date{\today}%
\pacs{05.40.-a}{Fluctuation phenomena, random processes, noise, and Brownian motion
}%
\pacs{87.16.Nn}{Motor proteins (myosin, kinesin dynein)
}%
\pacs{07.10.Cm}{Micromechanical devices and systems
}%
\abstract{
Particles moving in oscillating potential with broken mirror symmetry
are considered. We calculate their energetic efficiency, when acting
as molecular motors carrying a load against external force. It is shown
that interaction between particles enhances the efficiency in wide
range of parameters. Possible consequences for artificial molecular
motors are discussed.   
}
\begin{document}
\maketitle%

Molecular motors
\cite{magnasco_93,schliwa_03,jul_adj_pro_97,rei_han_02,reimann_02,han_mar_nor_04,lip_cha_klu_lie_mul_06,kol_fis_07}
are fascinating objects of study for nanoscience, be
it inspired by genuine biological questions of transport within the
cell or by the perspective of their artificial
manufacturing en masse
\cite{lin_hum_lof_99,mat_mul_03,ket_rei_han_mul_00,han_mar_08}.  

Indeed, in a living cell, many macromolecules move with the
help of some machine, usually made of various proteins and powered by
 ATP hydrolysis \cite{as_bi_96}. Not only
muscle contraction, bacterial flagellum movement and changes in the
shapes of amoebae are directly powered by the action of molecular
motors, but many more processes
\cite{schliwa_03,ras_kob_mal_fid_mas_04,del_benz_slu_aze_gol_06}
rely on motor proteins too.

 A very
challenging problem is the reported high 
efficiency of energy conversion in biological motors \cite{kol_fis_07}. 
%
%Several
%paradigmatic models are classically used in physics literature, mostly
%relying on the idea of rectification of Brownian motion into ordered
%particle current. Hence the name Brownian motors
%\cite{rei_han_02,reimann_02}, 
%which is sometimes
%used interchageably with the notion of molecular motors, although in
%our opinion there are some basic differences between the two terms. 
%
%One of the important points in this difference 
%touches the very question of energetic
%efficiency
%\cite{sekimoto_97,kam_hon_tak_98,par_dec_02,qian_04,par_bla_cao_bri_98}. 
%
On the model level, the energetic
efficiency was thoroughly studied 
\cite{sekimoto_97,kam_hon_tak_98,par_dec_02,qian_04,par_bla_cao_bri_98,sch_sei_08}.
It turns out that it is very low in classical setups of
Brownian motors, which is either the flashing or the rocking ratchet
\cite{reimann_02,par_dec_02}.  
On the other hand, a scheme named ``reversible ratchet''
\cite{parrondo_98,par_bla_cao_bri_98,ast_der_99} was
introduced, which was shown to be much more efficient, and high
efficiency was also characteristic of the two-state model \cite{jul_pro_95}.

%At the same
%time it cannot be appropriately considered as driven fully by thermal
%fluctuations, as is the case in proper Brownian motors. The signature if this
%fact is that the motor-induced current is non-zero even at zero
%temperature, when all thermal noise dies out. (We neglect the
%possibility of rectification of quantum fluctuations, which would be a
%completely different story \cite{han_mar_nor_04}.) 

One of the astonishing and not yet fully understood aspects of
molecular motors is their collective behaviour. Motor proteins carrying a
load very often act 
in small but coordinated groups  \cite{kol_fis_07}. In gene transcription and  
translation large number of motor proteins move along the same track
\cite{ras_kob_mal_fid_mas_04}. Strong interactions of hard-core type
between individual motors occur in such situations.

The collective movement of coupled Brownian motors was studied
\cite{rei_kaw_bro_han_99,stu_kol_06} 
as well as cooperative effects of
hard-rod molecules in thermal ratchets
\cite{der_vic_95,der_ajd_96,agh_men_pli_99}. 
In these studies, very complex dependence of
the current on the size of the molecules was established. 
Collective effects were found to induce
non-zero current even in mirror symmetric potential due to dynamical
symmetry breaking \cite{jul_pro_95}. Recent studies show that 
cooperative movement of several motors is the
generic feature of cargo transport within the cell 
\cite{klu_lip_05,lip_cha_klu_lie_mul_06}.

Slightly different perspective was adopted in the studies of
interactions in the transport of kinesin \cite{gre_gar_nis_sch_cho_07},
ribosomes  \cite{bas_cho_07}, and RNA polymerase
\cite{tri_cho_08}. Here, no details on the ratchet mechanism are
assumed and the processes at work are idealised in the set of
transition rates between the states representing spatial positions and
internal conformations of the motor proteins. 

In our work we assume that the motor particles move in explicit,
although crudely simplified, time-dependent potential of the type
 used in ``reversible ratchet'' models 
\cite{parrondo_98,par_bla_cao_bri_98}. We shall aim at
clarification of the role of interactions on the energetic efficiency
of the motors.

\begin{figure}[t]
\begin{center}
\includegraphics[scale=0.5]{%
\slaninafigdir/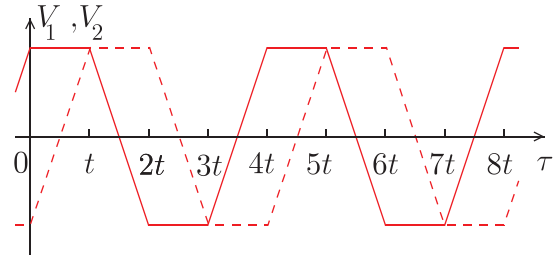}
\end{center}
\caption{%time-sequence....
{
Graph of the time dependence of the potential in which the particles
move. Full line: $V_1(\tau)$, dashed line: $V_2(\tau)$.
 The potential at the third site, $V_0$, is time independent.
}
} 
\label{fig:time-sequence}
\end{figure}

Our model consists of 
$N$ particles occupying integer positions on the segment of length
$L$, with periodic boundary conditions.  Let us denote $x_i(\tau)$ the
position of $i$-th  particle at time $\tau=1,2,3\ldots$. The
configuration of other particles as seen by  the $j$-th particle
 is described by the function
$n_j(x,\tau)=\sum_{i,i\ne j}\delta(x-x_i(\tau))$, where $\delta(x)=1$ if $x=0$
and zero otherwise.

The $j$-th particle moves
in the potential
\begin{equation}
U_j(x,\tau)=V(x,\tau)+x\,F+g\,n_j(x,\tau)
\label{eq:potential}
\end{equation}
composed of three parts.
The first term  the 
is spatially and temporally periodic external driving
potential $V(x,\tau)=V_{x\mathrm{~mod~}3}(\tau)$. 
We chose the smallest non-trivial spatial period 
$3$ for
convenience, although larger periods may offer further interesting
effects. The second part comes from the uniform and static external 
force $F$ against
which a useful work is done. Third, there is repulsive on-site interaction between
particles, with strength $g\ge 0$.

The time-dependent potential evolves in a four-stroke cycle 
{
sketched
in Fig. \ref{fig:time-sequence}.
}   
The duration of every  stroke is the quarter-period
 $t$.  The potential has three independent values,
$V_a(\tau)$, $a=0,1,2$, with $V_a(\tau)=V_a(\tau-4t)$.
We fix $V_0(\tau)=0$ and the other two  evolve in a 
piecewise
linear pattern 
\begin{equation}
\begin{split}
V_1&(\tau)=-V_1(\tau+2t)=\\
&=\left\{
\begin{array}{rlrrrr}
&V&\mathrm{~for~}&0&<\tau<&t\\
&V+2\,(1-\tau/t)\,V&\mathrm{~for~}&t&<\tau<&2t
\end{array}
\right.
\end{split}
\end{equation}
with a phase shift $V_1(\tau)=V_2(\tau+t)$.
Note, however, that the time $\tau$ is discrete, so the potential actually
undergoes finite jumps, rather than smooth change.
In the following we shall always set the amplitude to $V=1$. 
{
The
potential can be understood as a travelling wave. A simplistic picture
is that the particle
is captured within one of the minima and is drifted by the moving
wave. However, random diffusion, which is effective even at zero
temperature, makes the things complex.
}

\begin{figure}[t]
\includegraphics[scale=0.85]{%
\slaninafigdir/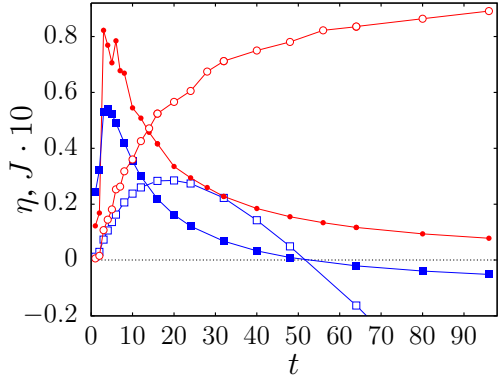}
\caption{%eff-curr-vs-speed-1200-1200-all-x-1-2....
Efficiency (empty symbols) and current (full symbols) as a function of
the quarter-period. The temperature is $T=0$ ({\Large $\circ$},
{\Large $\bullet$}) and $T=40$ ($\Box$, $\blacksquare$). Other parameters are
$L=1200$, $N=1200$, $g=1/9$,  and $F=0.2$.} 
\label{fig:eff-curr-vs-speed-1200-1200-all-x-1-2}
\end{figure}

\begin{figure}[t]
\includegraphics[scale=0.85]{%
\slaninafigdir/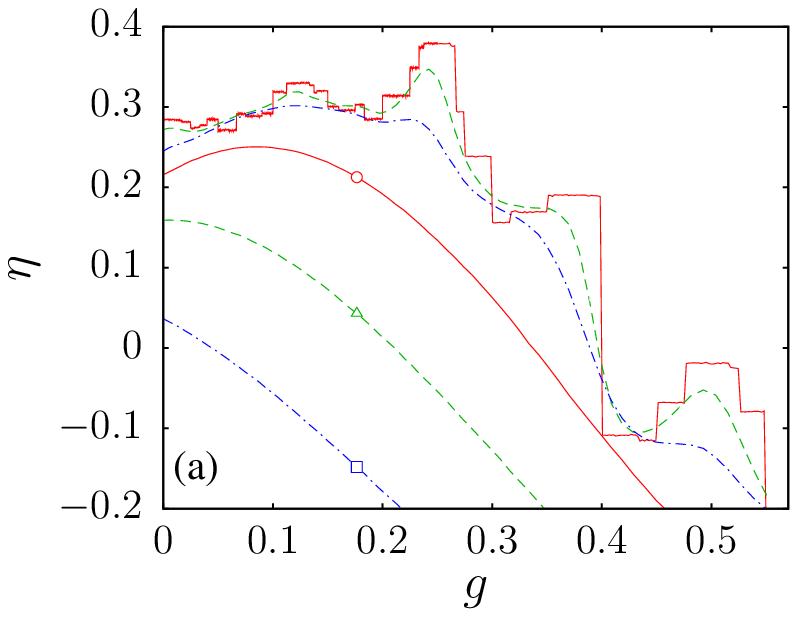}
\includegraphics[scale=0.85]{%
\slaninafigdir/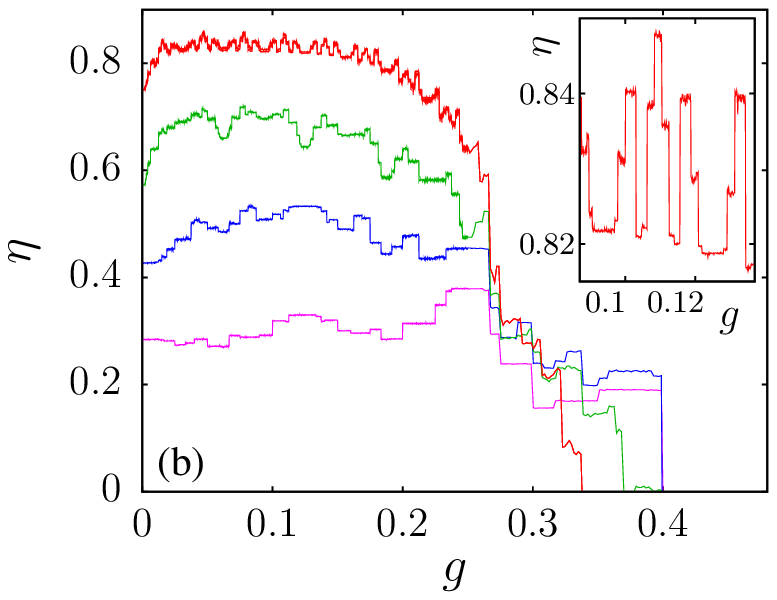}
\caption{%eff-vs-int-1200-1200-all-8-x-2....eff-vs-int-1200-1200-0-all-x-2....
Dependence of the efficiency on interaction strength, for $L=1200$,
$N=1200$, and $F=0.2$.   
Panel (a):
$t=8$,  temperatures
$T=0$ (solid line), $T=5$ (dashed line), $T=10$ (dash-dotted line),  
$T=30$ (solid line marked by {\Large $\circ$}), $T=60$ (dashed line
marked by $\bigtriangleup$), and $T=100$ (dash-dotted line marked by
$\Box$). 
Panel (b): $T=0$ and quarter-periods (from top to
bottom) $t=64$, $t=32$,  $t=16$, $t=8$,  and $t=4$. The ``noise'' in the
curve for $t=64$ is in fact a complicated deterministic dependence, as
can be seen in the inset, where enlarged particular of the curve is
shown.
}
\label{fig:eff-vs-int-1200-1200-all-8-x-2+eff-vs-int-1200-1200-0-all-x-2}
\end{figure}

The particles can move to nearest neighbour sites with probabilities
determined by the difference in potential (\ref{eq:potential}). Thus,
if $|x-y|=1$, the $j$-th  particle hops from site $x$ to $y$ with
probability
\begin{equation}
W_{j,x\to y}=\frac{1}{2}\,
\Big(1+\mathrm{e}^{\beta\,(U_j(y,\tau)-U_j(x,\tau))}\Big)^{-1}\;.
\label{eq:hopping-probability}
\end{equation}
For convenience we introduce the temperature as $T=270/\beta$.
{(The number $270$ is fairly arbitrary and was chosen with the aim
to see interesting things at ``aesthetic'' values of $T$.)}
At each  time step $\tau$ we select $N$ times a particle
randomly  and move it according to the probability
(\ref{eq:hopping-probability}). Therefore, each 
particle may move more than once in one time step and  when the number of particles is
very high, the probability 
that it moves $k$ times approaches the Poisson distribution with unit
average, $P(k)=1/(\mathrm{e}\,k!)$. This seemingly minor point plays important role in
analytical calculations we shall present later. 
{
It is also the source
of finite-size effects, briefly discussed at the end of this paper.
}
\begin{figure}[t]
\includegraphics[scale=0.85]{%
\slaninafigdir/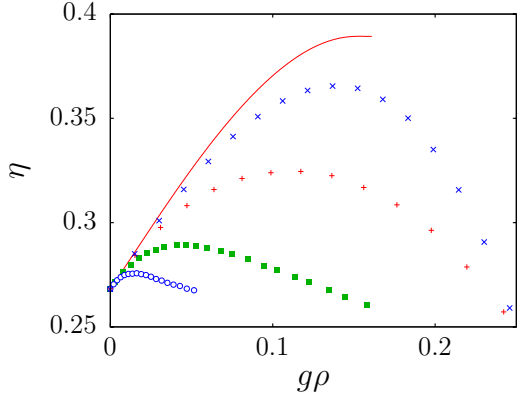}
\caption{%eff-vs-int-all-all-30-16-x-1....
Dependence of the efficiency on the interaction. To
facilitate the comparison of data at various densities with the
mean-field approximation, the product $g\rho$ 
is used as independent variable.  We plot the data for four densities,
$\rho=0.1$ ({\Large $\circ$}),   
$0.3$ ($\blacksquare$), $1$ ($+$), and $10$ ($\times$). The other
parameters are 
$F=0.1$, $T=30$, and $t=16$.
The solid line
is the result of the mean-field approximation.  
} 
\label{fig:eff-vs-int-all-all-30-16-x-1}
\end{figure}

The simplest quantity of interest is the current
\begin{equation}
J(\tau)=\sum_{i}\big(x_i(\tau+1)-x_i(\tau)\big)
\end{equation}
or rather its time average per particle
$J=\lim_{\tau\to\infty}(\tau\,N)^{-1}\,\sum_{\tau'=1}^\tau J(\tau')$. The main
focus of this work being on the energetics of the motor, we must
define the energy input and the useful work done. The latter quantity, at
time $\tau$, is $w(\tau)=F\,J(\tau)$. The
 energy pumped into the
motor from outside relates to the change of the potential
$V_a(\tau)$  while the particles stay immobile. Thus, the absorbed
energy between steps $\tau - 1$ and $\tau$ is
\begin{equation}
a(\tau) = \sum_i\big(V(x_i(\tau),\tau)-V(x_i(\tau),\tau-1)\big)
\end{equation}
and the efficiency, accordingly, 
\begin{equation}
\eta=
\frac{\lim_{\tau\to\infty}\sum_{\tau'=1}^\tau
  w(\tau')
}{
\lim_{\tau\to\infty}\sum_{\tau'=1}^\tau a(\tau')}\;.
\end{equation}

The typical results for the current and efficiency can be seen in
Fig. \ref{fig:eff-curr-vs-speed-1200-1200-all-x-1-2}. Generically,
both quantities decrease with increasing temperature.
Only for very fast driving i. e. at the
smallest values of $t$, the maxima of  current and efficiency 
occur at non-zero temperatures.

On the other hand, it is the region of larger $t$ that is interesting,
as the efficiency grows when $t$ gets longer. At the same time, the
current diminishes. This is easy to
understand, because slower driving implies lower current but also 
it leaves the system closer to
equilibrium and the loss of energy by non-equilibrium dissipation
is therefore also lower. However, as also seen from
Fig. \ref{fig:eff-curr-vs-speed-1200-1200-all-x-1-2}, 
for large enough temperature $T$ and
force $F$ the trend reverses and $\eta$ may be even negative. The
source of this behaviour is the decrease of the current, not
overweighted by decrease in absorbed energy, when driving
gets slower. If a non-zero external force $F$ is applied, the current
eventually changes sign and so does also the efficiency.

In
 Fig.
 \ref{fig:eff-vs-int-1200-1200-all-8-x-2+eff-vs-int-1200-1200-0-all-x-2}a
 we can see how the
efficiency depends on interaction strength. At zero temperature we
observe complicated sequence of steps. They are gradually smeared out
when temperature increases, until it reaches a function with single
maximum, which moves to lower $g$ and eventually disappears when
temperature is further increased. This leads to our main observation
that within certain interval of temperature (and other parameters),
efficiency can be enhanced by the interaction of the motors. However,
our simulations show that at the same time the current is suppressed. 
\begin{figure}[t]
\includegraphics[scale=0.85]{%
\slaninafigdir/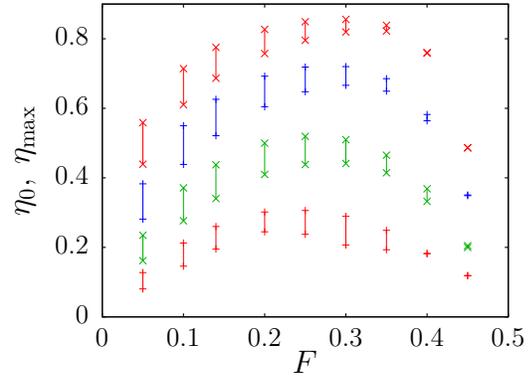}
\caption{%maxeff-vs-force-1200-1200-10-all-max-x....
Comparison of the efficiency at zero interaction $\eta_0$ with the
maximum efficiency $\eta_\mathrm{max}$ obtained by appropriately
tuning the interaction strength $g$. Each pair of points connected by
a vertical line denotes $\eta_0$ (lower point) and
$\eta_\mathrm{max}$ (upper point). Where only a single point is shown,
the maximum
is reached at $g=0$.  For all points, we have $L=N=1200$, $T=10$. The
sets of pairs correspond to quarter-periods $t=8$ (lower $+$), $16$
(lower $\times$),
$32$ (upper $+$), and $64$ (upper $\times$).
} 
\label{fig:maxeff-vs-force-1200-1200-10-all-max-x}
\end{figure}
In Fig
 \ref{fig:eff-vs-int-1200-1200-all-8-x-2+eff-vs-int-1200-1200-0-all-x-2}b
we show 
that the efficiency of the motor at $T=0$ exhibits
complicated dependence on interaction with lots of maxima and the
complexity keeps growing when the driving gets slower. Similar complex
pattern is seen also for the  current.

In Fig. \ref{fig:eff-vs-int-all-all-30-16-x-1} we can see the
typical dependence of the efficiency on interaction at a medium 
temperature. At higher density of
particles $\rho=N/L$ 
the maximum possible increase of efficiency with respect to
non-interacting case is higher. This is due to larger role of
fluctuations in local particle density, when the number of particles
is lower. We can see that the gain in efficiency can be easily larger
than $10\%$, a quite appreciable difference. The dependence of the
gain on external force and on the speed of driving is shown in Fig. 
\ref{fig:maxeff-vs-force-1200-1200-10-all-max-x}. We can see that the
efficiency reaches maximum at certain value of the force and beyond
that value the efficiency not only diminishes but eventually becomes
non-optimizable by tuning $g$. i. e. non-interacting case is the most
efficient. However, there is quite wide window, where the efficiency
can be tuned to optimum by setting both force and interaction to
convenient values.

\begin{figure}[t]
\includegraphics[scale=0.85]{%
\slaninafigdir/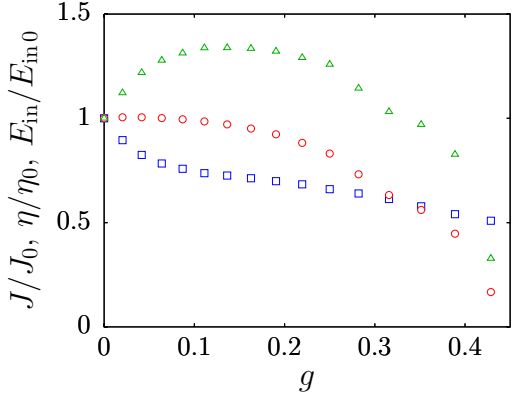}
\caption{%eff-curr-ener-vs-int-10-16-x-1....
{Current ({\Large $\circ$}), efficiency ($\bigtriangleup$) and energy
input ($\Box$) relative to their values for
$g=0$. The parameters are $L=N=1200$, $T=10$, $t=16$, and $F=0.1$.
}
}
\label{fig:eff-curr-ener-vs-int-10-16-x-1}
\end{figure}

\begin{figure}[t]
\includegraphics[scale=0.85]{%
\slaninafigdir/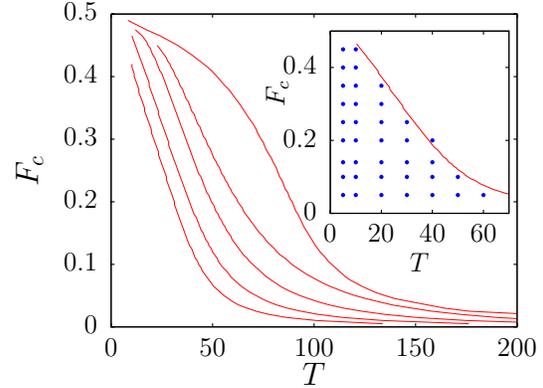}
\caption{%phasdiag-force-vs-temp-matlab...
Phase diagram of the ``reversible ratchet'', computed using the mean-field
approximation. The lines separate the
optimizable (lower-left) phase from the non-optimizable
(upper-right) and correspond to quarter-periods $t=1$, $8$, $16$,
$32$, and $64$ (from top to bottom). In the inset, the mean-field
phase diagram for $t=32$ 
accompanied by the results of numerical simulation for $\rho=0.3$. 
Each heavy dot in
the rectangular grid denotes a pair of variables $(T,F)$ for which
maximum efficiency was found at interaction strength $g>0$. 
} 
\label{fig:phasdiag-force-vs-temp-matlab}
\end{figure}

{
To gain further insight into the mechanism of efficiency enhancement,
we plot in Fig. \ref{fig:eff-curr-ener-vs-int-10-16-x-1} the
dependence of 
efficiency, work performed and energy input on interaction strength $g$, all relative 
to the value for non-interacting ($g=0$) case. We can see that for
small $g$, the work done is practically interaction-independent and
starts to decrease only at larger $g$. This behaviour copies the
dependence of the current, as work is simply current times force. On
the other hand, the energy input drops quickly even at small $g$. This
accounts for the net increase of efficiency at small $g$. For larger
$g$, the decrease in current (i. e. work), outweights the decrease of
input energy and the efficiency drops again. Obviously, both current
and efficiency are zero at exactly the same value of $g$. 

The reason why the input energy drops while current stays practically
constant when we switch on the interaction can be stated as follows. 
The particles are essentially driven by the travelling potential wave. At not
too high temperature, the depth of the minima of the potential does not
influence much the current. Now, weak interaction effectively shifts
the potential slightly upwards at the minima, while at the maxima
(where particles rarely occur) the potential is unchanged. The depth
of the travelling valleys is lowered, but the current carried by them
is practically the same. On the other hand, the energy is pumped in by
the temporal change of the potential at the position of the
particle. The effect of the repulsive 
interaction is that the particles are less
concentrated at the actual potential minimum, but reside with
relatively higher probability on the left or right to the minimum. Therefore,
the number of particles whose potential energy is increased at a
particular moment is lowered in comparison with the non-interacting
case. Hence the decrease in the value of the input energy. Note,
however, that this explanation is rather sketchy and more subtle
effects are also present here.
}

We complemented the numerical simulations by analytical calculations
using the mean-field approximation. 
Although the process representing the movement of the motors is
inherently non-stationary and principally non-Markovian, 
due to periodic external driving, we can describe it equivalently as a
Markov process on larger state space and find stationary state for that
process. If the
particles do not interact, it is enough to study the movement of a single
representative particle in the potential (\ref{eq:potential}). The probabilities
(\ref{eq:hopping-probability}) for one hop of the particle from site
$x$ to site $y$ form a time-dependent matrix $W_1(\tau)$. However, as
we mentioned, the number of hops the particle actually makes follows
the Poisson distribution. The actual transition matrix for the
movement of the particle at time $\tau$ is therefore{
$W(\tau)=\sum_{k=0}^\infty \frac{1}{\mathrm{e}\,k!} \big(W_1(\tau)\big)^k =
 \exp(W_1(\tau)-1)$}.
 Combined  positions of the
particle at times $\tau$, $\tau+1$, \ldots $\tau+4t-1$  evolve
according 
to a Markov process. The only non-zero
elements of the corresponding transition matrix  $W^R$ are
 $W^R_{y\,\tau+1,x\,\tau}=W_{yx}(\tau)$.
{
While $W(\tau)$ is a time-dependent $3\times 3$ matrix, $W^R$ is a
time-independent matrix $12t\times 12t$. 
If we divide it into $3\times 3$ blocks, we can see that only the
blocks below the block diagonal and the single block in the right top corner
are non-zero. These non-zero blocks contain the matrices $W(1),W(2),\ldots,W(4t)$.
}

The stationary state $p^R$ of this composite process is the solution
of 
the linear equation 
\begin{equation}
p^R=W^R\,p^R\;.
\label{eq:for-stationary-state}
\end{equation}
 The average
  current, work and absorbed energy are then found as linear
  combinations of the elements of  $p^R$. 

The interaction can be taken into account using the mean-field
approximation  
$g\,n_j(x,\tau)\simeq g\,\rho\,p^R_{x\,\tau}$. Through the hopping
probabilities (\ref{eq:hopping-probability}) the stationary state
$p^R$ penetrates into the matrix  $W^R$ and the equation (\ref{eq:for-stationary-state})
becomes non-linear. 

We can judge from Fig. \ref{fig:eff-vs-int-all-all-30-16-x-1} to what
extent the mean-field approximation (MFA) captures the results obtained by
numerical simulations. We observe that MFA tends to coincide with
simulations when the average density is large, more precisely in the
limit $\rho\to\infty$ with $g\rho$ kept fixed. 
{The reason why MFA works well in the high-density
limit (and not in the opposite case $\rho\to 0$, as one might think)
is that MFA is based on neglecting the relative density fluctuations. 
When the average
density is low, changing position of a single particle implies large
relative change of local density, while in the high-density regime the
change in local density is very small. 
}

It also suggests that
the slope of the dependence of $\eta$ on $g\rho$ for $g\rho\to 0$ is reproduced 
exactly in MFA.
If the slope is positive, the efficiency can be increased  
by turning on the interaction. We shall call
that optimizable phase. Negative slope
means that non-interacting case is the best possible and we shall call
that phase non-optimizable. From
Fig. \ref{fig:eff-vs-int-all-all-30-16-x-1} we can also see that the
sign of the slope does not depend on the
density $\rho$. Therefore, we can draw a phase diagram in the
temperature-force plane, like in
Fig. \ref{fig:phasdiag-force-vs-temp-matlab}. The line determines the temperature
dependence of the critical force $F_c$ where the slope vanishes.
 We can see that the area of the optimizable phase shrinks
when the driving gets slower ($t$ increases). This is easy to
understand, as slower driving itself induces higher efficiency, so
there is less room left for further optimization by tuning the
interaction strength. In Fig. \ref{fig:phasdiag-force-vs-temp-matlab}
we also demonstrate that the simulation results for the phase diagram
agree reasonably well with the MFA even for relatively low density
$\rho=0.3$, where otherwise the MFA is ill-justified, as seen from
Fig. \ref{fig:eff-vs-int-all-all-30-16-x-1}. 

\begin{figure}[t]
\includegraphics[scale=0.85]{%
\slaninafigdir/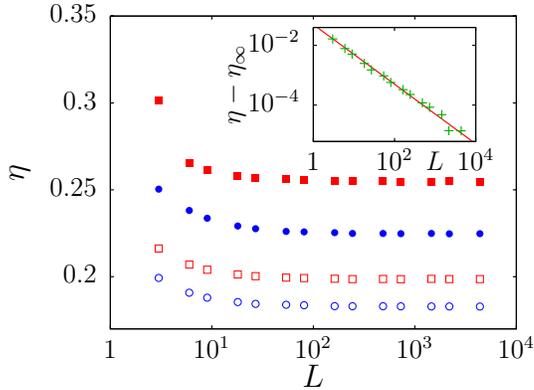}
\caption{%eff-vs-size-30-8-0-14....
{Effect of finite size on the efficiency, for fixed density $\rho=1$.
Various symbols correspond to $T=5$ (squares), $T=30$ (circles), $g=0$
(empty symbols), $g=0.1$ (filled symbols). 
Further parameters are $t=8$, $F=0.14$. In the inset, deviation of
the efficiency from its asymptotic ($L\to\infty$) value, for $\rho=1$,
$T=30$, $t=8$, $g=0$ and $F=0.14$. The line is the power $\propto L^{-1}$. 
}
}
\label{fig:eff-vs-size-30-8-0-14}
\end{figure}

{
We also investigated the issue of finite-size effects. In Fig. \ref{fig:eff-vs-size-30-8-0-14}
we show the efficiency for various widths $L$ of the sample, from
$L=3$ to $4374$ (recall that $L$ must be a multiple of $3$), with fixed
density. We can see that the finite size effects do occur, but can be
neglected for sizes larger than about $L\simeq 100$. We can also see
that the deviation of the size-dependent efficiency $\eta$ from its
asymptotic value $\eta_\infty$ for $L\to\infty$ scales like
$\eta-\eta_\infty\sim L^{-1}$. The finite-size effects
stem from the probability that a chosen particle makes exactly $k$
moves in one step of the simulation. The distribution converges to
Poissonian in the limit $L\to\infty$, but definitely deviates from
it if the number of particles is finite.
}

To summarise, within the scheme of ``reversible ratchet'', we showed 
that interaction of large number of molecular motors leads to
collective effects resulting in substantial increase of energetic
efficiency. The price to pay is the decrease of particle current. We
also found that the dependence of the efficiency (and current) on
interaction strength is very complex at low temperatures, exhibiting
complex sequence of multiple maxima and minima. For certain values of
the external force it
 can even become a sequence of current reversals. To
check the generic character of our results we performed similar
simulations also for the case of rocking ratchet
\cite{bar_han_kis_94}. Also here, we observed increase of efficiency in a window
of temperatures and external forces. However, the overall efficiency
is order of magnitude lower than in the ``reversible ratchet'', thus
also the increase is much lower in absolute numbers.

The question to discuss is how we can tune the interaction strength in
a real system, where the interaction of motors has actually a hard-core
character. In fact, although the motivation of our study was to large
extent based on biological experiments, we had in mind possible
applications in artificial motors \cite{han_mar_08} like those of 
Refs. \cite{mat_mul_03,ket_rei_han_mul_00}, made of micropores. 
Varying the spatial
geometry of the pore we can effectively tune the influence of
hard-core repulsion of moving particles. Thus we have another degree
of freedom at our disposal in our design of efficient man-made molecular 
motors.

\acknowledgments
I gladly acknowledge inspiring discussions with P. Chvosta, E. Ben-Jacob and P. Kalinay. 
This work was carried out within the project AVOZ10100520 of the Academy of 
Sciences of the Czech republic and was 
supported by the Grant Agency of the Czech Republic, grant no. 
202/07/0404.

\end{document}